\documentclass[twocolumn,letter]{jpsj3}

\usepackage{amsmath}
\usepackage{amsfonts}
\usepackage{amssymb}
\usepackage{cite}
\usepackage{txfonts}
\usepackage{bm}
\usepackage{tabularx}
\usepackage{graphicx,color}

\def\Hc2{H_\mathrm{c2}}
\def\Tc{T_\mathrm{c}}

\author{
	Shunichiro \textsc{Kittaka}$^{1,\thanks{E-mail: kittaka@issp.u-tokyo.ac.jp}}$, 
	Toshiro \textsc{Sakakibara}$^{1}$, 
	Masato \textsc{Hedo}$^{2}$, 
	Yoshichika \textsc{\={O}nuki}$^{2,3}$, and
	Kazushige \textsc{Machida}$^{4}$
}

\inst{$^{1}$Institute for Solid State Physics, University of Tokyo, Kashiwa, Chiba 277-8581, Japan\\
      $^{2}$Department of Physics, University of the Ryukyus, Nishihara, Okinawa 903-0213, Japan\\
      $^{3}$Department of Physics, Osaka University, Toyonaka, Osaka 560-0043, Japan\\
      $^{4}$Department of Physics, Okayama University, Okayama 700-8530, Japan
}

\begin{document}

\title{Verification of Anisotropic $s$-Wave Superconducting Gap Structure in CeRu$_2$ \\from Low-Temperature Field-Angle-Resolved Specific Heat Measurements}

\date{\today}

\abst{
The field-angle-resolved specific heat $C(T,H, \phi)$ of the $f$-electron superconductor CeRu$_2$ ($\Tc=6.3$~K) has been measured at low temperatures down to 90~mK on two single crystals of slightly different qualities.
We reveal that the $C(\phi)$ oscillation in a rotating magnetic field, originating from the gap anisotropy, 
diminishes at low temperatures below the characteristic field $H^\ast$, 
as expected for an anisotropic gap without nodes.
We also observe the suppression of $H^\ast$ by decreasing the gap anisotropy ratio $\Delta_{\min}/\Delta_{\max}$,
a behavior that has been predicted from a microscopic theory for anisotropic $s$-wave superconductors. 
The present technique is established as a powerful tool for investigating minimum-gap structures as well as nodal structures.
}

\kword{superconducting gap, anisotropic $s$-wave, CeRu$_2$, field-angle-resolved specific heat}

\maketitle

The determination of superconducting (SC) gap structures is one of the important subjects for characterizing superconductivity and resolving its pairing mechanism. 
Field-angle-resolved specific heat $C(T,H,\phi)$ and thermal conductivity $\kappa(T,H,\phi)$ measurements are being established 
as powerful tools for identifying the position of nodes in the SC gap.\cite{Sakakibara2007JPSJ,Matsuda2006JPCM}
In the so-called "Doppler-shift" regime, microscopic theories~\cite{Vekhter1999PRB,Miranovic2003PRB,Miranovic2005JPC,Vorontsov2006PRL,Hiragi2010JPSJ}
predict that the zero-energy (ZE) density of states (DOS), $N_0$, in nodal superconductors oscillates under a rotating magnetic field $H$ and becomes minimum 
when $H$ is parallel to the nodal directions. 
Indeed, the nodal structures of CeCoIn$_5$ and CeIrIn$_5$ have been identified to be of the $d_{x^2-y^2}$-wave type 
by probing the field-orientation dependence of DOS through $C(\phi)$ and $\kappa(\phi)$ under a rotating $H$.\cite{An2010PRL,Izawa2001PRL,Kittaka2012PRB,Kasahara2008PRL}

Note that not only the position of nodes but also that of gap minima can be clarified using the same technique:
ZE DOS becomes minimum when $H$ is applied along the minimum-gap direction.
Moreover, this technique has been proposed to be effective in evaluating the gap anisotropy ratio $\Delta_{\min}/\Delta_{\max}$.
According to the calculations based on a microscopic theory, 
the ratio of the ZE DOS maximum to the ZE DOS minimum under a rotating $H$ in the limits $T \rightarrow 0$ and $H \rightarrow 0$ remains larger than unity for a nodal gap, 
whereas it becomes unity for an anisotropic full gap.\cite{Miranovic2005JPC}
This is because, in the case of a full gap, quasiparticles (QPs) are confined to each vortex core in the low-$T$ and low-$H$ regime,
leading to the isotropic field-orientation dependence of ZE DOS.
By increasing $H$ above the characteristic field $H^\ast$, the spatial extensions of ZE DOS from all cores extend and overlap with each other, and 
the gap anisotropy contributes to the field-orientation dependence of DOS. 
Because ZE QPs with a larger energy gap are more strongly confined in vortex cores, 
$H^\ast$ tends to become smaller as the minimum-gap size $\Delta_{\rm min}$ decreases (in the nodal-gap case, $H^\ast \sim 0$).
Nevertheless, experimental evidence of these theoretical predictions has not been obtained yet,
probably because a low-$T$ ($T < 0.1\Tc$) experiment is essential.

The cubic Laves phase compound CeRu$_2$ is an ideal material for examining the theoretical predictions for anisotropic $s$-wave superconductors.
It exhibits superconductivity with a relatively high transition temperature $\Tc=6.3$~K and an upper critical field $\mu_0\Hc2=5.2$~T.
Although $C(T)$ exhibits a conventional-BCS-type $T$ dependence, $C(H)$ shows a nearly-$\sqrt{H}$ behavior at 0.5~K ($\sim 0.08 \Tc$), \cite{Hedo1998JPSJ}
implying the occurrence of low-energy QP excitations without nodes.
In addition, a coherence peak appears in the nuclear-spin lattice relaxation rate $1/T_1$ of $^{101}$Ru by doping a small amount of Al into the Ru site,
although this feature is strongly suppressed in high-quality samples.\cite{Mukuda1998JPSJ}
These observations offer strong evidence of an anisotropic $s$-wave gap without nodes.
The gap anisotropy ratio $\Delta_{\min}/\Delta_{\max}$ has been estimated to be 0.74 from the NMR measurements on powder samples \cite{Mukuda1998JPSJ}
and 0.45 from the angle-integrated photoemission spectroscopy on high-quality single crystals.\cite{Kiss2005PRL}

In a previous study,\cite{Yamada2007JPSJ}
the $C(\phi)$ of a single-crystalline CeRu$_2$ was measured under $H$ rotated around the $[001]$ axis in the field range of 0.5~T $\le \mu_0H \le 6$~T and 
in the temperature range of 0.3~K $\le T \le 2$~K.
Under these conditions, a clear fourfold $C(\phi)$ oscillation was observed with its minimum in $H \parallel \langle 110 \rangle$, 
indicating the presence of gap minima in the $\langle 110 \rangle$ direction.
Moreover, at 0.5~T ($\sim 0.1\Hc2$), 
the normalized oscillation amplitude $A_4$ decreases toward zero upon cooling below about 1~K.
This feature can be explained qualitatively by a microscopic theory assuming an anisotropic $s$-wave gap with $\Delta_{\min}/\Delta_{\max}=1/3$.
Nevertheless, the key feature for distinguishing between nodes and gap minima, i.e., $A_4 \rightarrow 0$ with $T \rightarrow 0$, has not yet been fully established 
because $A_4$ was still finite at the lowest temperature of 0.3~K.
In this study, we have investigated the $C(T,H,\phi)$ of two CeRu$_2$ samples down to 90~mK and
evidenced that $A_4$ indeed becomes zero in the low-$T$ and low-$H$ regime.
Furthermore, we have provided experimental evidence of the interrelation between $H^\ast/\Hc2$ and $\Delta_{\min}/\Delta_{\max}$.

\begin{figure}
\begin{center}
\includegraphics[width=3.3in]{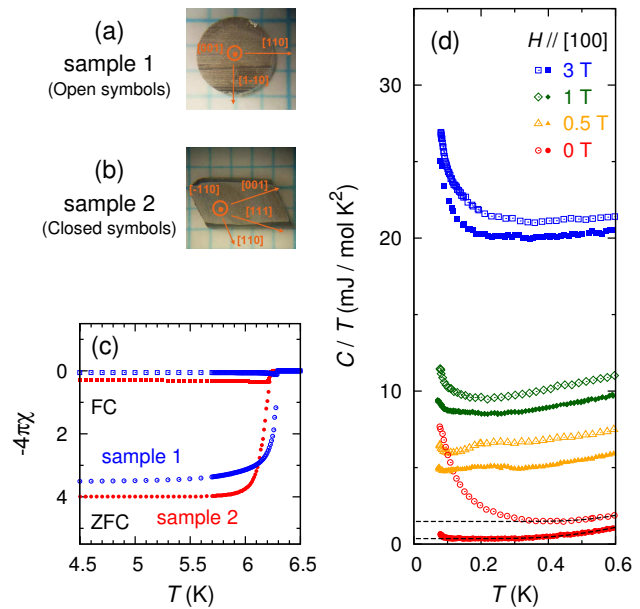}
\end{center}
\caption{
(Color online) (a), (b) Pictures of samples 1 and 2. 
(c)~Temperature dependence of the dc susceptibility $\chi=M/H$ normalized by $-1/4\pi$ 
measured at 1~mT applied along the $[001]$ and $[\bar{1}10]$ axes for samples 1 (open) and 2 (closed).
The data taken in the zero-field-cooling (ZFC, circles) and the field-cooling (FC, squares) processes are shown.
(d)~Temperature dependence of the specific heat divided by temperature, $C/T$, measured at $\mu_0H=0$, 0.5, 1, and 3~T applied parallel to the $[100]$ axis. 
Open and closed symbols represent the results for samples 1 and 2, respectively.
}
\label{Tdep}
\end{figure}

Two clean single crystals of CeRu$_2$ (samples 1 and 2) grown by the Czochralski pulling method were used in the present study.
Sample~1 is identical to the crystal used in a previous $C(\phi)$ study~\cite{Yamada2007JPSJ}, which is in the shape of a disk with flat surfaces parallel to the (001) plane (114 mg weight).
Sample~2 is in the shape of a plate with its flat surfaces parallel to the $(\bar{1}10)$ plane (144 mg weight).
The pictures of both samples are shown in Figs.~\ref{Tdep}(a) and \ref{Tdep}(b).
The specific heat $C$ was measured by a standard quasi-adiabatic heat-pulse method in a dilution refrigerator (Oxford Kelvinox AST Minisorb).
In all the data presented, the addenda contribution, which hardly depends on the field angle, was subtracted.
Magnetic fields were generated in the $xz$ plane using a vector magnet composed of horizontal split-pair (5~T) and vertical solenoidal (3~T) coils.
By using a stepper motor mounted at the top of a magnet Dewar, we rotate the refrigerator around the $z$ axis.
Thus, the entire system enables us to control the magnetic field orientation three-dimensionally.
For each field angle, the specific heat was determined by the average of approximately seven successive measurements.
To investigate the $\Tc$ of each sample, 
magnetization measurements were performed using a commercial magnetometer (Quantum Design, model MPMS) under a magnetic field of 1~mT applied perpendicular to the flat surfaces.


Figure \ref{Tdep}(c) shows the $T$ dependence of the dc susceptibility $\chi=M/H$ normalized by the ideal full-Meissner value, $-1/4\pi$.
The data of samples 1 and 2 are represented by open and closed symbols, respectively.
Owing to the large demagnetization factor of the samples, $-4\pi\chi$ exceeds 1 in the full-Meissner state.
A sharp SC transition is observed at $\Tc=6.28$~K for sample~1 and at 6.22~K for sample~2. 
Here, $\Tc$ is defined as the temperature at which the zero-field-cooling susceptibility reaches 10\% of the full Meissner signal. 
The sharpness of the SC transition and the high $\Tc$ ensure the high quality of both samples.

Low-temperature specific heat measurements, however, indicate the inclusion of a low concentration of impurities in sample~1.
Figure \ref{Tdep}(d) shows the $T$ dependence of $C/T$ for $T \le 0.1\Tc$ investigated at several fields applied parallel to the $[100]$ axis.
Here, the open and closed symbols represent the data obtained using samples~1 and 2, respectively.
At zero field, the $C/T$ of sample~1 exhibits a large upturn below 0.3~K.
This upturn is strongly suppressed by applying a weak magnetic field of 0.5~T, the behavior that cannot be explained by the nuclear specific heat contribution.
In sharp contrast, no prominent upturn is present in the $C/T$ of sample~2. 
These facts imply the presence of impurities in sample~1.
Although the $\Tc$ of sample~1 is slightly higher than that of sample~2,
it has been reported that a low concentration of impurities in Ce sites can enhance $\Tc$ in CeRu$_2$.\cite{Mukuda1998JPSJ}

The dashed lines in Fig.~\ref{Tdep}(d) represent the fits to the $C(T)$ data at zero field
using the BCS formula $C(T) \sim a\exp(-\Delta/k_{\rm B}T)+\gamma_{\rm res}T$.
Here, $a$ is a constant, $\Delta$ the SC gap amplitude, $k_{\rm B}$ the Boltzmann constant, and $\gamma_{\rm res}$ the residual value of $C/T$.
The fitting range was chosen to be 0.35~K $\le T \le 0.6$~K for sample~1 and 0.14~K $\le T \le 0.6$~K for sample~2.
In such a low-$T$ range, the parameter $\Delta$ gives a rough estimate of $\Delta_{\rm min}$.
From the fitting, we obtain $\gamma_{\rm res}=1.48$~mJ/(mol K$^2$) and $\Delta_{\rm min}=4.2$~K for sample 1, and $\gamma_{\rm res}=0.35$~mJ/(mol K$^2$) and $\Delta_{\rm min}=2.4$~K for sample 2.
The larger $\Delta_{\rm min}$ in sample~1 also supports the presence of impurities 
because magnetic impurity scatterings tend to average the gap anisotropy in anisotropic $s$-wave superconductors.

When the BCS fit to $C(T)$ is carried out in the intermediate-$T$ range of 1.5~K $\le T \le 4.2$~K, $\Delta$ of 11.7~K has been obtained.\cite{Hedo1998JPSJ}
Because, in such a high-$T$ range, QPs are thermally excited anywhere across the gap, 
this $\Delta$ gives a rough estimate of the average amplitude of the gap.
Hence, $\Delta_{\rm min}/\Delta_{\rm max}$ is estimated to be at most $1/3$ for sample~1 and $1/5$ for sample~2.
This $\Delta_{\rm min}/\Delta_{\rm max}$ of sample~1 coincides with that 
used for the calculations of the ($T$, $H$) map of $A_4$ for the same sample in a previous $C(\phi)$ study.~\cite{Yamada2007JPSJ}

At fields above 1~T, the $C(T)$ of both samples exhibits a Schottky-type anomaly below 0.2~K.
This contribution at 3~T is comparable to the specific heat of $^{99}$Ru and $^{101}$Ru nuclei ($I=5/2$) at a magnetic field 
calculated using the electric quadrupole interaction parameters obtained by NQR experiments.\cite{Matsuda1995JPSJ,Mukuda1998JPSJ}.
At lower fields, however, the observed Schottky anomaly is much smaller than what would be expected for the nuclear contribution.
This apparent inconsistency arises due to a long $T_1$; 
at zero field, it exceeds 100 s below 1~K. \cite{Mukuda1998JPSJ}
For such a long $T_1$, the nuclear spins are decoupled and cannot be detected by specific heat measurements with a measuring time of several tens of seconds. 
For convenience, we have subtracted only the nuclear Zeeman contribution $C_{\rm n} \propto H^2/T^2$ for the data presented below.

\begin{figure}
\begin{center}
\includegraphics[width=3.3in]{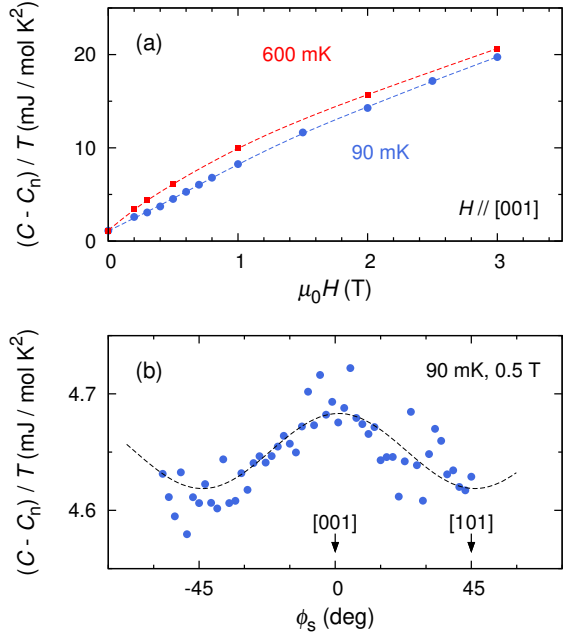}
\end{center}
\caption{
(Color online) 
(a)~Field dependence of $(C-C_{\rm n})/T$ for sample~2 in $H \parallel [001]$ measured at 90 and 600~mK.
(b)~Variation in $(C-C_{\rm n})/T$ for sample~2 at a magnetic field of 0.5~T rotated around the $[010]$ axis measured at 90~mK.
}
\label{Hdep}
\end{figure}

Figure \ref{Hdep}(a) shows the $H$ dependence of $(C-C_{\rm n})/T$ for sample~2 in the low-$H$ regime for $H \parallel [001]$ obtained at 600 and 90~mK.
At 600~mK, $C(H)$ exhibits $\sqrt{H}$-like behavior, consistent with the previous report.\cite{Hedo1998JPSJ}
By decreasing $T$ down to 90~mK, 
the initial slope of $C(H)$ is suppressed and the $H$-linear dependence becomes prominent.
This behavior is reminiscent of the features predicted for anisotropic $s$-wave superconductors,\cite{Nakai2004PRB,Nakai2006PRB}
although it apparently contradicts the relatively small $\Delta_{\rm min}/\Delta_{\rm max}$ ($\sim 1/5$) of sample~2.

Now, let us turn our attention to the field-orientation dependence of $C/T$.
Figure~\ref{Hdep}(b) shows the $C(\phi_{\rm s})$ of sample~2 measured 
at $H$ of 0.5~T rotated around the $[010]$ axis at 90~mK,
where $\phi_{\rm s}$ is the angle between $H$ and the $[001]$ axis in the $(010)$ plane. 
Note that the $(010)$ and $(001)$ planes are equivalent in cubic structure. 
As shown in Fig.~\ref{Hdep}(b), 
a large fourfold oscillation with its minimum along the $\langle 110 \rangle$ axes is observed.
The dashed line in Fig.~\ref{Hdep}(b) represents the fitting result obtained using the expression $C(T,H,\phi_{\rm s}) - C_{\rm n}(T,H) = C_0(T) + C_H (T,H)(1+A_4(T,H)\cos 4\phi_{\rm s})$, 
where $C_0$ and $C_H$ are the zero-field and field-dependent components, respectively, and $A_4$ is the relative amplitude of the fourfold oscillation.
Interestingly, the $A_4$ of sample~2 at 0.5~T is as high as 1\% even at a low temperature of 90~mK, although it is at most 0.3\% for sample~1 at 0.3~K.\cite{Yamada2007JPSJ}
This oscillation cannot be attributed to the nuclear contribution because the normal-state $C(\phi)$ measured under a sufficiently high magnetic field of 6~T does not show any oscillation at 0.32~K \cite{Yamada2007JPSJ}.

\begin{figure}
\begin{center}
\includegraphics[width=3.3in]{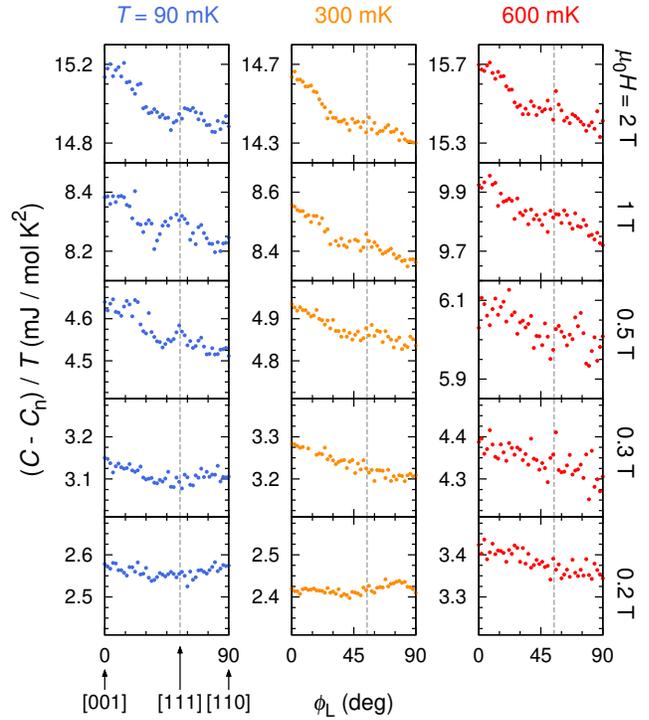}
\end{center}
\caption{
(Color online) 
Variation in $(C-C_{\rm n})/T$ for sample~2 in a magnetic field rotated around the $[\bar{1}10]$ axis.
The normalized amplitude $A_4$ of the fourfold $C(\phi_s)$ oscillation is estimated using the relation $C(\phi_{\rm L}=0^\circ)-C(90^\circ)=2C_HA_4$.
}
\label{phi-dep}
\end{figure}

To get more insight into the field-orientation dependence,
we measured $C(\phi_{\rm L})$ at various temperatures and fields under a rotating $H$ within the $(\bar{1}10)$ plane that includes the [001], [111], and [110] directions.
Here, $\phi_{\rm L}$ denotes the angle between $H$ and the $[001]$ axis in the $(\bar{1}10)$ plane.
As shown in Fig.~\ref{phi-dep}, 
$C(\phi_{\rm L})$ in the high-$H$ regime shows a maximum at $\phi_{\rm L}=0$~deg and a minimum at $\phi_{\rm L}=90$~deg.
This anisotropy in $C(\phi_{\rm L})$ is consistent with that in $C(\phi_{\rm s})$,
indicating the presence of a gap minimum in the $\langle 110 \rangle$ directions.
Furthermore, a small peak is seen in $C(\phi_{\rm L})$ at $\phi_{\rm L} \simeq 56$~deg, 
suggesting the presence of a local gap maximum in the $\langle 111 \rangle$ directions. 
This gap maximum is naturally formed when gap minima exist in the cubic $\langle 110 \rangle$ directions.

\begin{figure}
\begin{center}
\includegraphics[width=3.3in]{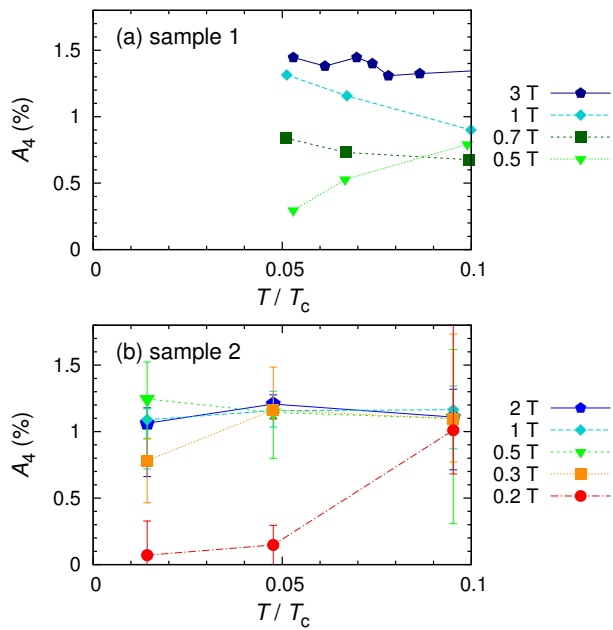}
\end{center}
\caption{
(Color online) 
Normalized fourfold amplitude $A_4$ as a function of $T/\Tc$ ($\Tc=6.3$~K) at several magnetic fields for (a) sample~1 \cite{Yamada2007JPSJ} and (b) sample~2.
The broken lines are guides to the eye.
}
\label{A4}
\end{figure}

\begin{figure}
\begin{center}
\includegraphics[width=3.3in]{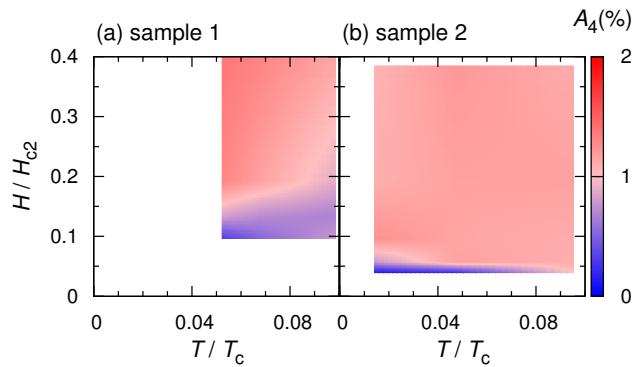}
\end{center}
\caption{
(Color online) 
Contour plots of $A_4(T,H)$ for (a) sample 1 and (b) sample 2 using the data in Figs.~\ref{A4}(a) and \ref{A4}(b), respectively. 
Here, $\Tc=6.3$~K and $\mu_0\Hc2=5.2$~T.
}
\label{contour}
\end{figure}

Using the relation $C(\phi_{\rm L}=0^\circ)-C(90^\circ)=2C_HA_4$,
we evaluate the $A_4(T,H)$ of sample~2 from the data in Fig.~\ref{phi-dep}. 
The obtained $T$ dependence of $A_4$ for sample~2 is plotted in Fig.~\ref{A4}(b).
By decreasing $H$ down to 0.2~T, the $A_4$ of sample~2 is suppressed to be nearly zero below 0.3~K;
no sign change nor increase in $A_4$ has been observed down to 90~mK ($\sim 1/70 \Tc$).
Note that the temperature of 90~mK is $\sim 1/30$ of $\Delta_{\rm min}$ ($\sim 2.4$~K);
QP excitations are strongly suppressed anywhere in the gap, as in the isotropic-full-gap case ($A_4=0$). 
For comparison, the $A_4(T)$ of sample~1\cite{Yamada2007JPSJ} is also plotted in Fig.~\ref{A4}(a),
although the data below 0.3~K was not available due to the upturn in $C/T$ attributed to the impurity effect.
Because $A_4$ is not constant but field dependent at finite fields, the $C(\phi)$ anisotropy cannot be attributed to the localized QPs that contribute to $C \propto H$.
Figures~\ref{contour}(a) and \ref{contour}(b) show the contour plots of $A_4(T,H)$ using the experimental data in Figs.~\ref{A4}(a) and \ref{A4}(b).
We empirically define the characteristic magnetic field $H_{\rm exp}^\ast$ by the field where $A_4$ is suppressed to be less than 0.5\% at low $T$. 
From our experimental results, the $H_{\rm exp}^\ast$ values are 0.5~T ($\sim 0.1\Hc2$) for sample~1 and 0.2~T ($\sim 0.05\Hc2$) for sample~2.

On theoretical grounds, 
$A_4$ at low $T$ is expected to approach zero with decreasing field below $H^\ast$ for anisotropic $s$-wave superconductors, and 
$H^\ast/\Hc2$ is inferred to become small with decreasing $\Delta_{\rm min}/\Delta_{\rm max}$. \cite{Miranovic2005JPC,Yamada2007JPSJ}
Recall that $\Delta_{\rm min}/\Delta_{\rm max}$ of sample~2 estimated from the BCS fit to the $C(T)$ data is roughly half of that of sample~1.
This fact proves that $H^\ast/\Hc2$ is, indeed, closely related to $\Delta_{\rm min}/\Delta_{\rm max}$.
In addition, we do observe $A_4 \simeq 0$ for sample~2 in the low-$T$ and low-$H$ regime.
Thus, the significant features of anisotropic $s$-wave superconductors theoretically proposed have been verified in the present experiment.

In summary, we have investigated the field-angle-resolved specific heat of two single crystals of CeRu$_2$ with slightly different qualities.
From the temperature dependence of the specific heat, the gap anisotropy ratio $\Delta_{\rm min}/\Delta_{\rm max}$ is estimated to be $1/3$ and $1/5$ for samples~1 and 2, respectively.
We found that the amplitude of the $C(\phi)$ oscillation under a rotating magnetic field is suppressed to be almost zero at low temperatures 
when the field is lower than the characteristic field $H^\ast$ (0.5~T for sample~1 and 0.2~T for sample~2).
These features are in good agreement with the theoretical predictions for anisotropic $s$-wave superconductors: 
the absence of the $C(\phi)$ oscillation in the limits $T \rightarrow 0$ and $H \rightarrow 0$ and the suppression of $H^\ast$ by decreasing $\Delta_{\rm min}/\Delta_{\rm max}$.
The present results demonstrate that the field-angle-resolved measurement is indeed a valid technique for investigating the anisotropic superconducting gap structures. 
A detailed investigation of the relationship between the gap anisotropy and the Fermi surface topology would pave the way 
to understanding the microscopic mechanism of gap minima in CeRu$_2$.

We acknowledge K. Kitagawa, Y. Aoki, H. Sato, and K. Izawa for helpful discussions.
K. M. thanks the Aspen Center for Physics for hospitality during the summer workshop (NSF Grant No. 1066293).
This work has been partially supported by Grants-in Aid for Scientific Research on Innovative Areas ``Heavy Electrons'' (20102007, 23102705)
and ``Topological Quantum Phenomena'' (25103716) from MEXT, and KAKENHI (25800186, 21340103) from JSPS.


\end{document}